\documentclass[aip,reprint,amsmath,amssymb,floatfix]{revtex4-1}

\usepackage{graphicx,xcolor,times}
\usepackage{bm,mathptmx,etoolbox}

\begin{document}
\title{Atomistic Framework for Glassy Polymer Viscoelasticity Across Twenty Frequency Decades}

\author{Ankit Singh}
\email{ankit.singh@unimi.it}
\affiliation{Department of Physics ``A. Pontremoli'', University of Milan, Via Celoria 16, 20133 Milan, Italy}

\author{Vinay Vaibhav}
\email{vinay.vaibhav@uni-goettingen.de}
\affiliation{Department of Physics ``A. Pontremoli'', University of Milan, Via Celoria 16, 20133 Milan, Italy}
\affiliation{Institut für Theoretische Physik, University of Göttingen,
Friedrich-Hund-Platz 1, 37077 Göttingen, Germany}

\author{Caterina Czibula}
\email{caterina.czibula@tugraz.at}
\affiliation{Institute of Bioproducts and Paper Technology, Graz University of Technology, Inffeldgasse 23, 8010 Graz, Austria}

\author{Astrid Macher}
\affiliation{Polymer Competence Center Leoben GmbH, Sauraugasse 1, 8700 Leoben, Austria}

\author{Petra Christ{\"o}fl}
\affiliation{Polymer Competence Center Leoben GmbH, Sauraugasse 1, 8700 Leoben, Austria}

\author{Karin Bartl}
\affiliation{Institute for Chemistry and Technology of Materials, Graz University of Technology, Stremayrgasse 9, 8010 Graz, Austria}

\author{Gregor Trimmel}
\affiliation{Institute for Chemistry and Technology of Materials, Graz University of Technology, Stremayrgasse 9, 8010 Graz, Austria}

\author{Timothy W. Sirk}
\affiliation{US Army DEVCOM Army Research Laboratory, Aberdeen Proving Ground, Maryland 21005, United States}

\author{Alessio Zaccone}
\email{alessio.zaccone@unimi.it}
\affiliation{Department of Physics ``A. Pontremoli'', University of Milan, Via Celoria 16, 20133 Milan, Italy}

\begin{abstract} 
Glassy polymers are central to engineering applications, yet their viscoelastic response over broad frequency and temperature ranges remains difficult to characterize. We extend non-affine deformation theory by incorporating a time-dependent memory kernel within the Generalized Langevin Equation for atomistic non-affine motions, yielding frequency-dependent mechanical response. Applied to poly(methyl methacrylate) (PMMA), the method captures the shear modulus and relaxation spectrum across more than twenty decades in frequency, from hundreds of terahertz to the millihertz regime, thus bridging polymer mechanics from ordinary to extreme scales. Our predictions show quantitative consistency with independent estimates from oscillatory-shear molecular dynamics, Brillouin scattering, ultrasonic spectroscopy, Split-Hopkinson testing, and dynamic mechanical analysis (DMA), demonstrating a unified theoretical–computational route for multiscale characterization of polymer glasses.
\end{abstract}

\maketitle

%%%%%%%%%%%%%%%%%%%%%%%%%%%%%%%%%%%%%%%%%%%%%%%%%%%%%%%%%%%%%%%%%%%%%%%%%%%%%%%%%%
Glassy polymers are essential in structural and protective applications, yet their viscoelastic response across experimentally relevant time and frequency scales remains difficult to describe consistently \cite{colmenero_2015polymers, wang_2024glassy}. High-frequency molecular simulations (GHz–THz) provide insight into relaxation dynamics near the glass transition, but low-frequency behavior (Hz) relevant for load-bearing applications \cite{astm_2013standard,lakes2009viscoelastic,lakes_2017viscoelastic} are typically modeled through phenomenological frameworks \cite{tschoegl2012phenomenological} that compress broad relaxation spectra into effective parameters. As a result, establishing a direct molecular-to-macroscopic link in polymer-glass mechanics remains a central challenge.

Experimentally, distinct techniques probe disjoint frequency windows: DMA accesses mHz–Hz \cite{menard2020dynamic, MULLERPABEL_2022_107701,HENRIQUES_2018_394,Zaccone_book, Wyss_PhysRevLett.98.238303, Sun_PhysRevB.111.184107, MULLIKEN20061331}; Split-Hopkinson tests reach the kHz regime \cite{MULLIKEN20061331, Sahraoui_1994, moy_paul_2003}; ultrasonic spectroscopy spans MHz \cite{afifi2003ultrasonic, afifi2003annealing}; and Brillouin scattering extends into the GHz regime \cite{Merklein_2022, weishaupt1995pressure, prevot_2025_softmatter}, where recent work shows consistency with atomistic modeling \cite{delasoudas2025elasticity}. Although these methods reveal different segments of the relaxation spectrum, they do not independently provide a unified microscopic picture.

On the computational side, elastic moduli of disordered solids have been explored through stress-fluctuation approaches and related formalisms \cite{Barrat_1988, Leonforte_PhysRevB.72.224206, Procaccia_PhysRevE.93.063003, kriuchevskyi_2017numerical, Krief_PhysRevE.103.063307,theodorou,lempesis_2013tracking,Doros, wittmer_2013shear, Wittmer_PhysRevE.91.022107, Kriuchevskyi_PhysRevLett.119.147802}, activation–relaxation theories \cite{Dyre_PhysRevB.53.2171, Olsen_PhysRevLett.81.1031}, parallel-replica dynamics \cite{perez_2015parallel}, and non-affine approaches \cite{Leonforte_silica, Zaccone_book, Scossa, vaibhav2024time, singh2025viscosity}. However, direct MD remains limited to a narrow, high-frequency window due to computational constraints \cite{mukherji}.

Non-affine lattice dynamics (NALD) offers a microscopic route to connect atomic structure and macroscopic elasticity \cite{elder, vaibhav2024time}. In NALD, structural disorder generates unbalanced affine forces that drive non-affine relaxations \cite{Zaccone_book}, producing a negative correction to the Born modulus \cite{Scossa}. The vibrational density of states and affine force correlator together determine the frequency-dependent modulus. However, previous formulations typically assumed a constant (Markovian) memory kernel, whereas glassy polymers exhibit broad, non-Markovian relaxation spectra requiring memory-dependent friction \cite{milster_2024tracer, li2017computing}.

Here we extend NALD by introducing a power-law memory kernel within the Generalized Langevin Equation, enabling the description of secondary relaxations. We apply this framework to PMMA, a well-studied polymer glass \cite{MULLIKEN20061331, richeton_2007_modeling, ionita_2015viscoelastic, hu_2016_experimental, Lu_1997_polymer, William_PhysRevApplied.18.064078}, using atomistic simulations to compute the vibrational modes \cite{wales_2006potential, keyes1994unstable, Laird, stratt1995instantaneous, Jack, vaibhav2025experimental} and affine force correlator. The resulting shear modulus spans more than twenty decades in frequency, from terahertz vibrational dynamics to millihertz mechanical response, and agrees with oscillatory-shear MD, Brillouin spectroscopy \cite{weishaupt1995pressure}, ultrasonic data \cite{afifi2003ultrasonic}, Split-Hopkinson tests \cite{MULLIKEN20061331, Sahraoui_1994, moy_paul_2003}, and our DMA measurements, consistent with the known presence of local relaxations \cite{Ngai, MULLIKEN20061331, moy_paul_2003, Sahraoui_1994}. Supported by analytic low-frequency scaling of $\Gamma(\omega)$ and $g(\omega)$ \cite{Scossa, Milkus_2017,palyulin2018parameter}, the method bridges optical and mechanical regimes and reveals the full multiscale relaxation spectrum of PMMA, with implications for glass-transition modeling \cite{glova}. A key result of this work is a description of polymer viscoelasticity across timescales of conventional mechanical testing (DMA), high rate tests (Split-Hopkinson testing, ultrasonics, and Brillouin spectroscopy), and molecular dynamics, ultimately spanning twenty frequency decades across molecular processes to application-relevant timescales. 

For the numerical study, we have considered 64 polymeric chains (each containing 10 monomers) of PMMA in a periodic simulation box, making a system of $N = 9920$ atoms. Further details are provided in Section S2 of the Supplementary Information (SI) \cite{supp}. The atomic interactions are modeled using the General AMBER Force Field (GAFF), where bond and angle terms are harmonic and dihedral interactions are described by periodic potentials. Electrostatic interactions are treated using the particle-particle particle-mesh (PPPM) method, with a real-space cut-off of $9$ \AA. Lennard-Jones interactions are truncated at $9$ \AA. Large-scale molecular dynamics simulations are carried out using LAMMPS \cite{LAMMPS} to generate polymer configurations under fixed temperature and pressure conditions. The system temperature and pressure are controlled through the Nosé–Hoover thermostat and barostat, respectively, with the pressure maintained at zero throughout the simulation. The equations of motion are solved under periodic boundary conditions using the velocity-Verlet integrator with a timestep of 1~fs.

In NALD, the potential energy surface (PES) \cite{wales_2006potential} is required to measure the mechanical response of atomistic systems. PES describes the total potential energy of the system as a function of the atomic positions, effectively defining the energy landscape in which the atoms reside. The first derivative of the PES with respect to atomic positions gives the forces acting on each atom, which govern the atomic motion and local stability of the system, including internal stresses. The second derivative, known as the Hessian or dynamical matrix, provides information about the curvature of the PES, which determines the vibrational modes and eigenfrequencies of the system. These vibrational properties are crucial for calculating the affine (infinite-frequency) elastic modulus as well as the non-affine corrections arising from local, disorder-induced atomic rearrangements. We diagonalize the Hessian matrix to obtain the eigenfrequencies and corresponding eigenmodes of the system that are used to calculate the normalized vibrational density of states $g(\omega)=\frac{1}{3N-3}\sum_{i}\delta(\omega-\omega_{i})$ and the affine force field correlator $\Gamma_{\alpha\beta\kappa\chi}(\omega)=\langle\hat{\Xi}_{\alpha \beta} \hat{\Xi}_{\kappa\chi}\rangle_{\omega \in [\omega_{k},\omega_{k}+\delta\omega_{k}]}$ (see section S1 of SI \cite{supp} for definition), where $\hat{\Xi}_{\alpha \beta} (\omega)$ is the component of affine force field. The frequency-dependent elastic constants \( C_{\alpha\beta\kappa\chi}(\Omega) \) can be obtained in terms of these quantities via the relation \cite{Zaccone_book, Scossa}
\begin{align}
    C_{\alpha\beta\kappa\chi}(\Omega) = C_{\alpha\beta\kappa\chi}^{A}-3 \rho\int \frac{\Gamma_{\alpha\beta\kappa\chi}(\omega) g(\omega)}{m(\omega^{2}-\Omega^{2})+i\Omega{\nu}} d\omega,
    \label{Cint}
\end{align}
\noindent
where \( \Omega \) is the applied strain frequency, \( C^{\mathrm{A}}_{\alpha\beta\kappa\chi} \) is the affine contribution to elasticity, $\rho$ is number density, and $\nu $ is the friction memory kernel. The integrand term accounts for non-affine effects that act as a negative correction to the quasi-static limit, arising from local asymmetries in the atomic potential energy landscape during macroscopically affine deformations, as typically observed in glasses as well as in noncentrosymmetric crystals \cite{cui}.

The various components of the elastic constants determined using Eq.~(\ref{Cint}) depend on different components of the affine force field with the corresponding component of the affine elastic constant. Here, we focus on the shear modulus \( G \), for which Eq.~(\ref{Cint}) can be rewritten in the form 
\begin{align}
    G^{*}(\Omega) = G_{A}-3 \rho \int \frac{\Gamma(\omega) g(\omega)}{(\omega^{2}-\Omega^{2})+i\Omega{\nu}} d\omega.
    \label{Gint}
\end{align}
Here, \( G^{*}(\Omega) \) represents the complex shear modulus and  $G_{A}$ is the affine modulus. We separate the real and imaginary components of \( G^{*}(\Omega) \) (SI \cite{supp}, section S1) that correspond to the storage and loss moduli, respectively, which quantify the elastic and viscous responses of the system. We also absorb the mass term in the affine force correlator and friction memory kernel \cite{kriuchevskyi2020scaling, vaibhav2024time} to get corresponding mass-scaled quantity. 

For simplicity, a fixed value of memory kernel \( {\nu} \) is generally used in Eq.~\eqref{Gint}, which is the only parameter adjusted to perfectly match the MD results for the infinite-frequency plateau, as done also in \cite{vaibhav2024time,elder} and discussed in Section S1 of SI \cite{supp}. In a Markovian process, friction is treated as a constant, as is typically assumed in molecular simulations using, for example, the Langevin thermostat. However, in real glassy materials, friction is more accurately described within a non-Markovian framework, where the memory kernel depends on time or the history of deformation. It is possible to calculate the memory kernel using either the velocity autocorrelation function (VACF) or the force autocorrelation function through a reconstruction technique \cite{koch_2024analysis,jung_2017iterative,li2017computing}. In polymeric systems, the memory kernel typically exhibits a power-law behavior at intermediate and long times \cite{powerlawmemory_2025}. To capture experimental conditions, we allow the friction parameter \( \nu \) to vary and adopt a simple power-law decay \cite{milster_2024tracer, li2017computing} in the time domain, ${\nu}(t) \sim \nu_{0} \,t^{\delta-1}$ with $\delta < 1 $. In this work we use the memory kernel in the frequency domain as the corresponding Fourier transform that gives ${\nu}(\Omega) = {\nu}_{0} \Omega^{-\delta}$, indicating that the effective friction decreases at high frequencies and recovers the constant Markovian limit as \( \delta \to 0 \). We use this memory kernel in the NALD equations to calculate the shear modulus in order to compare with the experimental results.

%%%%%%%%%%%%%%%%%%%%%%%%%%%%%%%%%%%
\begin{figure}[t]
    \includegraphics[width=0.97\linewidth]{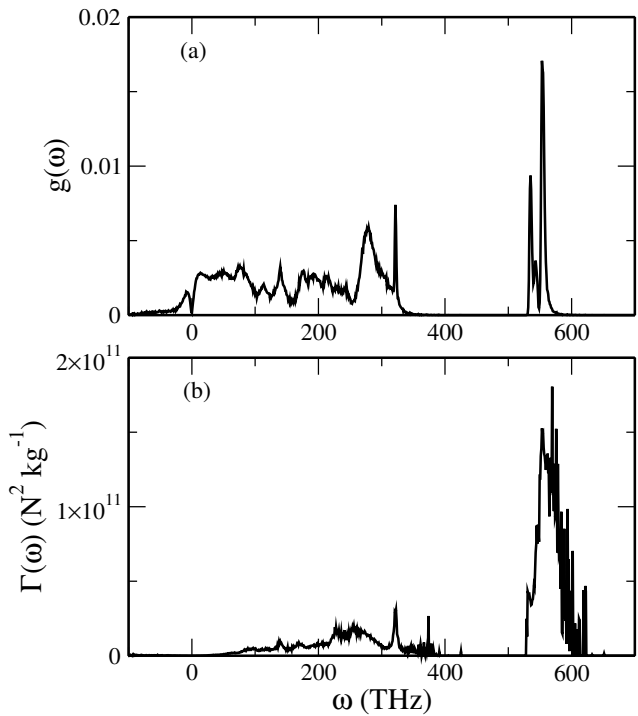}
    \caption{(a) Vibrational density of states $g(\omega)$ and (b) affine force field correlator $\Gamma(\omega)$ as a function of normal mode frequency $(\omega)$ at temperature $T=300$~K, for our simulated PMMA glass.}
    \label{fig1}
\end{figure}
%%%%%%%%%%%%%%%%%%%%%%%%%%%%%%%%%%%

We employ the NALD method to calculate the viscoelastic response and compare the results with experimental measurements. For the NALD calculations, we require microscopic information, namely the vibrational density of states \( g(\omega) \), the affine force field correlator \( \Gamma(\omega) \), and the friction memory kernel \( \nu \). In Fig.~\ref{fig1}(a) and (b), we plot the vibrational density of states (vDOS) and the affine force field correlator $\Gamma(\omega)$, respectively, as functions of the normal mode frequency $(\omega)$ at room temperature, $T=300$K. Since our system is at a finite temperature, the vDOS contains a significant fraction of imaginary modes (instantaneous normal modes, INMs) (conventionally represented on the negative branch of $\omega$) in addition to real modes \cite{keyes1994unstable, Laird, stratt1995instantaneous, Jack, vaibhav2025experimental}. The imaginary INMs originate from the presence of saddle points in the potential energy landscape, whereas the real modes correspond to stable vibrational motions (e.g. phonons).
The vDOS calculated from the Fourier transform of VACF \cite{mukherji_2024_vDOS_vacf} and as well as from neutron scattering experiments \cite{szymoniak_2025vDOS}, only provide the behaviour in the real-frequency domain.

We observe several peaks across the entire frequency spectrum. These features may arise from the characteristic bonding nature of PMMA, where certain bonding interactions contribute vibrational modes in the high-frequency range of the vDOS. Near zero frequency, on both the positive and negative branches, the vDOS increases linearly, as often observed in glasses and supercooled liquids \cite{Petry,Laird,palyulin2018parameter}. This low-frequency regime shows only little variation with temperature (as shown in Fig.~S1 of SI \cite{supp}).

In Fig.~\ref{fig1}(b), the affine-force field correlator \( \Gamma(\omega) \) is shown for the same temperature. The overall profile of \( \Gamma(\omega) \) exhibits several peaks, similar to the vDOS, with peak intensities increasing modestly as the temperature decreases (SI \cite{supp}, section S1). At high frequencies, around \( 550 \,\mathrm{THz} \), a pronounced peak is observed; however, its influence on the shear modulus is minimal. In contrast, at low frequencies, \( \Gamma(\omega) \) follows the expected analytical scaling ~\cite{Scossa, Milkus_2017} $\Gamma(\omega) \sim \omega^2$, which plays a dominant role in the determination of the modulus. This low-frequency behavior reflects the long-wavelength vibrational modes that couple strongly with the macroscopic deformation response.

Using the information from the vibrational density of states \( g(\omega) \) and the affine force field correlator \( \Gamma(\omega) \), we calculate the shear modulus as a function of the external frequency \(\Omega\) at different temperatures below the glass transition. An important additional quantity in this calculation is the friction memory kernel, which enters the equations in the power-law form \cite{milster_2024tracer, li2017computing} $\nu(\Omega) = \nu_{0} \Omega^{-\delta}$ with parameters \(\nu_{0} = 5.2 \times 10^{18} \,\mathrm{kg\,s^{-1}}\) and $\delta=0.35$.  In the time domain, this form corresponds to the memory kernel given by \(\nu(t) \sim \nu_{0}  t^{(\delta-1)}\). We plot and discuss \(\nu\) in the SI \cite{supp} as Figure~S4 shows \(\nu(\Omega)\) as a function of the external frequency.

In the determination of the shear modulus $G$, we perform the integration over both the real and imaginary branches of \(\omega\), as shown in Eqs.~(S.4, S.5) of SI \cite{supp}. Since the system is relatively small, finite-size effects arise, leading to the presence of non-physical low-frequency modes. To address this, a fraction of the small-frequency modes below a cut-off frequency \(\omega_{\rm min}\) is discarded during the integration, as the system cannot support propagating modes with frequencies lower than the speed of sound in the medium. The minimum frequency can be estimated as $\omega_{\rm min} = \frac{2\pi}{L} \sqrt{\frac{G}{\rho}}$, where \(L\) is the size of the simulation box, \(\rho\) is the mass density, and \(G\) is the shear modulus \cite{vaibhav2024time}. Hence, when performing the finite analytical integration for the modulus calculations, we incorporate all modes with $|\omega| < \omega_{\min}$ using analytical theory, to account for the low-frequency contributions below the cut-off (cf. Fig. S2 in the SI \cite{supp}).

We observe a secondary $\beta$ relaxation \cite{Ngai} in the shear modulus in the frequency domain around 1~Hz. This is reflected in the time domain, where $G(t)$ decays on a time-scale $\tau \sim 1$ s. Capturing such small frequencies in simulations is very challenging due to system size limitations and the computational cost of Hessian diagonalization. To overcome this, we use an analytical-theory extension in the very low-frequency regime, exploiting the known behavior of $\Gamma(\omega)$ and $g(\omega)$ \cite{Zaccone_book}. Specifically, we fit $\Gamma(\omega) = a_{1}\,\omega^{2}, \quad g(\omega) = a_{0}\,\omega,$ with $a_{1} =3.1 \times 10^{-18}$ and $a_{0} = 4.5 \times 10^{-16}$ as shown in Figs.~S2 and S3 of the SI \cite{supp}. We then evaluate the additional analytical contribution, obtaining the exact integral solution at finite integration limits for both the real and imaginary parts of the eigenfrequencies. In this regime, the friction kernel is found to be on the order of $\nu = 5 \times 10^{22}\,\mathrm{kg\,s^{-1}}$ across the entire deformation frequency range.

%%%%%%%%%%%%%%%%%%%%%%%%%%%%%%%%%%%
\begin{figure}[t]
    \centering
    \includegraphics[width=0.998\linewidth,clip]{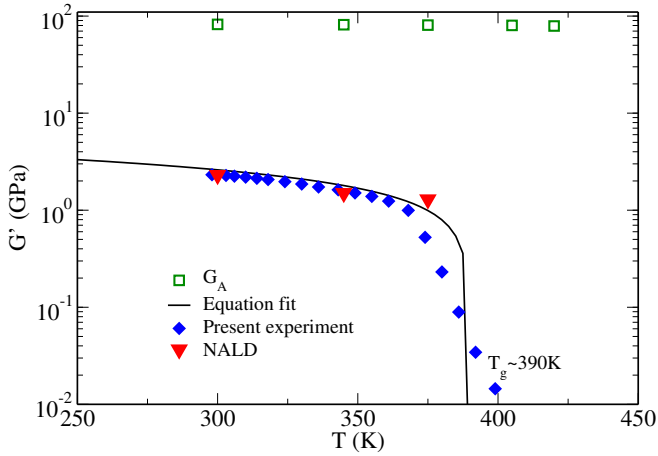}
    \caption{Temperature dependence of shear storage modulus \( G' \) for PMMA calculated using NALD (closed triangles) is compared with DMA experimental data (closed diamonds, at external frequency $\Omega=1$ Hz). The solid line represents the theoretical fit based on the equation described in SI \cite{supp}, with a glass transition temperature of \( T_g \approx 390~\mathrm{K} \). We also plot the affine modulus $G_{A}$ (open squares) as a function of temperature.}
    \label{fig2}
\end{figure}
%%%%%%%%%%%%%%%%%%%%%%%%%%%%%%%%%%%

In Fig.~\ref{fig2}, we show the shear modulus \(G'\) serving as an order parameter that distinguishes the glassy and liquid phases. Here, \(G'\) is plotted as a function of temperature at a very low deformation frequency of 1~Hz. We compare the experimental DMA results (closed diamonds) with atomistic NALD calculations (closed triangles), which show a reasonable agreement. We also fit the experimental data using the theoretical Zaccone-Terentjev expression for $G'$ vs $T$ (details in SI \cite{supp}) \cite{ZET}, capturing the sharp decay near the glass transition temperature \(T_g = 390~\mathrm{K}\)~\cite{PhysRevMaterials_tg, mohammadi_2017glass_tg, chimenti2023toward}. Although the NALD calculations capture the slow decrease in the modulus below $T_g$, the sharp decay near $T_g$ is not well captured. Moreover, we observed a slight decrease in the affine modulus $G_A$ near and above $T_g$ due to enhanced monomer mobility and reduced topological constraints.

%%%%%%%%%%%%%%%%%%%%%%%%%%%%%%%%%%%
\begin{figure}[t]
    \includegraphics[width=0.998\linewidth,clip]{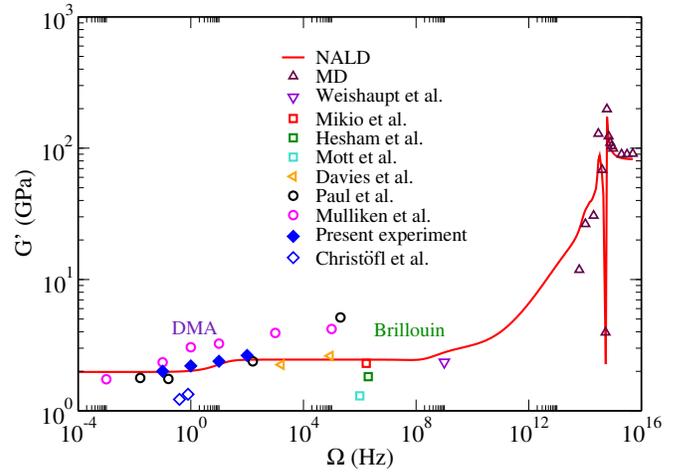}
    \caption{Low-frequency shear storage modulus $G^{\prime}$ as a function of external deformation frequency $\Omega$ at temperature $T=300$~K. The Solid line represents NALD calculations and symbols represent MD and experimental results that include DMA study in this work and other experiments from previous studies \cite{weishaupt1995pressure, Mikio_1995, afifi2003ultrasonic,afifi2003annealing, MOTT2008572, Sahraoui_1994,  davies1963dynamic, moy_paul_2003, MULLIKEN20061331, christofl2021comprehensive}.}
    \label{fig3}
\end{figure}
%%%%%%%%%%%%%%%%%%%%%%%%%%%%%%%%%%%

In Fig.~\ref{fig3}, we plot the shear modulus of PMMA as a function of the external frequency \(\Omega\) at \(T = 300\)~K, representing room temperature conditions. The results of the NALD atomistic calculation are compared with MD results, our experimental DMA data (details in section S3 of SI \cite{supp}) as well as literature data from different mechanical testing techniques, across a broad frequency range. Four distinct frequency regimes, which correspond to structural processes within PMMA, are observed and reliably captured by NALD. First, at the high-frequency end (above terahertz region), a plateau emerges corresponding to the affine modulus, a purely elastic contribution determined by the instantaneous bonding polymer chain or network; the second regime is the resonance frequency region (THz), where two sharp peaks appear: the first is at $300$~ THz, and the second one is close to $600$~THz, due to the vibrational resonances of PMMA chemical bonds \cite{ali_2015review}; the atomistic MD simulations accurately reproduce the resonance behavior. The third regime is the GHz regime, which coincides with experimental Brillouin spectroscopy results \cite{weishaupt1995pressure}. Here, non-affine deformations become significant. The first decay in modulus arises from non-affine relaxations that reduce the affine contribution to the shear modulus, resulting in mechanical softening \cite{MULLERPABEL_2022_107701, joaquin_2025aging}.  

At kHz frequencies, we include Split-Hopkinson test data \cite{MULLIKEN20061331, Sahraoui_1994, moy_paul_2003}, noting that the strain rate used to define the testing speed is not directly equivalent to a frequency; however, the stress wave generated during testing typically lies in the kHz range. Some experimental data agree well with NALD predictions, whereas others either overestimate or underestimate the modulus (For strain-rate and modulus conversions, see Section S4 of the SI \cite{supp}). Since Split-Hopkinson pressure bars are custom-built and testing conditions vary, such deviations are not unexpected. At lower frequencies, available techniques are limited to contact-based methods, each with its own limitations, leading to additional scatter. Literature shows that, even for a well-studied glassy polymer such as PMMA, an exact modulus is not possible to obtain from classical tests alone \cite{christofl2021comprehensive}. Measurements spanning nano- to macroscales demonstrate that local material variations and experimental constraints (e.g., deformation mode, sample preparation) only allow one to define a range of values, with nanoscale AFM experiments typically yielding lower moduli \cite{christofl2021comprehensive, ganser2018combining}. In our experimental study, we use PMMA from the same manufacturer for DMA experiments in the low-frequency regime (0.1–100 Hz) (details in section S3 of SI \cite{supp}). The DMA results match the NALD estimates and reveal a secondary decay in the shear modulus associated with the $\beta$-relaxation process \cite{MULLIKEN20061331, moy_paul_2003, Sahraoui_1994}.

To summarize, we provided the atomic-scale computation of the frequency-dependent shear modulus of a glassy polymer (PMMA) across a broad frequency spectrum, from several hundreds of terahertz down to millihertz. The computation scheme enables multiscale bridging in time, thanks to the calculation of non-affine atomic displacements, which contribute substantial negative corrections to the modulus, as the deformation frequency decreases. The calculations are successfully compared with MD simulations of the deformation process (limited to the terahertz regime), as well as with several sets of experimental data comprising both Brillouin light scattering (at GHz) and mechanical testing (from $10^5$ Hz down to millihertz). While a near-parameter-free quantitative match is possible, in order to capture the secondary $\beta$ relaxation in the modulus that occurs at hundreds of Hertz, a power-law memory function for the atomic-scale memory kernel is required, in agreement with previous results in the literature \cite{milster_2024tracer, li2017computing}. Overall, the proposed NALD framework allows us to bridge the timescale gap between non-destructive light scattering techniques and conventional mechanical testing of real soft materials, and provides a theoretical, experimentally validated description of the mechanics of glassy polymers across twenty orders of magnitude in deformation frequency, with profound implications for both fundamental science and engineering applications. A future challenge will be to extend this atomistic framework to the vicinity of the glass transition \cite{glova} and to more complex material architectures \cite{Varol}.

\section*{Data Availability}
The data that support the findings of this study are available within the article and its Supplementary Information.\\
%%%%%%%%%%%%%%%%%%%%%%%%%%%%%%%%%%%%%%%%%%%%%%%%%%%%%%%%%%%%%%%%%%%%%%%%%%%%%%%%%
\section*{Acknowledgements} AS, VV, and AZ gratefully acknowledge funding from the European Union through Horizon Europe ERC Grant number: 101043968 “Multimech”. VV acknowledges the computational resource provided via the project ``TPLAMECH'' on INDACO platform at the HPC facility of Università degli Studi di Milano. AZ gratefully acknowledges funding from the US Army DEVCOM Army Research Office through contract nr.  W911NF-22-2-0256. CC acknowledges the Hertha Firnberg program (project no. T 1314-N) of the Austrian Science Fund (FWF), Grant DOI: 10.55776/T1314 for funding. Discussions with Simone Napolitano and Federico Caporaletti are gratefully acknowledged.

%{\it Research contributions-}

\bibliographystyle{apsrev4-1}
\bibliography{article}
\end{document}